\documentclass[11pt,a4paper,twoside]{article}

\usepackage{amsmath}
\usepackage{amssymb}
\usepackage{amsfonts}
\usepackage{amsthm}
\usepackage{dsfont}

\newcommand{\BEQ}{\begin{eqnarray}}
\newcommand{\EEQ}{\end{eqnarray}}
\newcommand{\ket}[1]{\left | \, #1 \right \rangle}

\newcommand{\bra}[1]{\left \langle \, #1 \right |}
\renewcommand{\H}{\mathcal{H}}
\newcommand{\beq}{\begin{equation}}
\newcommand{\eeq}{\end{equation}}
\newcommand{\1}{\mathds{1}}

\begin{document}

{\bf \Large \begin{centerline}{
 The quantum world is not built up from correlations}
 \end{centerline}}
 
\begin{centerline}{ Found. Phys. {\bf 36}, 1573-1586 (2006).}
\end{centerline}
 \vskip1cm
{\large{\begin{centerline}{Michael Seevinck}\end{centerline}}
 \vskip1cm
{\large{\begin{centerline}{
Institute of History and Foundations of Science, Utrecht University,}
\end{centerline}\begin{centerline}{
P.O Box 80.000, 3508 TA Utrecht, The Netherlands.}\end{centerline}
\begin{centerline}{ E-mail: seevinck@phys.uu.nl}
\end{centerline}}
\vskip3cm
\noindent
It is known that the global state of a composite quantum system can be
completely determined by specifying correlations between measurements performed on subsystems only.
Despite the fact that the quantum correlations thus suffice to reconstruct the quantum state,
we show, using a Bell inequality argument, that they cannot be regarded as
objective local properties of the composite system in question.
It is well known since the work of
 J.S. Bell, that one cannot have locally preexistent values for all physical quantities,
  whether they are deterministic or stochastic.
 The Bell inequality argument we present here shows this is also impossible for
correlations among subsystems of an individual isolated
composite system.
Neither of them can be used to build up a world consisting of some
local realistic structure. As a corrolary to the result we argue that
entanglement cannot be considered ontologically robust.
 The argument has an important advantage over others because it
 does not need perfect correlations but only statistical correlations.
 It can therefore easily be tested in currently
 feasible experiments using four particle entanglement.
\vskip1cm
\noindent
\section{INTRODUCTION}\label{intro}
  What is quantum mechanics about? This question has haunted the physics community
ever since the conception of the theory in the 1920's. Since the
work of John Bell we know at least that quantum mechanics is not
about a local realistic structure built up out of values of
physical quantities \cite{bell}. This is because of the well known
fact that if one considers the values of physical quantities to
be locally real, they must obey a so-called Bell inequality, which
quantum mechanics violates. The paradigmatic example of a quantum
system giving rise to such a violation is the singlet state of two
spin-${\frac{1}{2}}$ particles. This state describes two particles that
are anti-correlated in spin. Bell's result shows that no single
particle in the singlet state can be regarded to have a locally
preexistent spin value. Instead, the
singlet state tells us that upon measurement the spin values, if
measured in the same direction on each particle, will always be
found anti-parallel. Because this (anti-) correlation is found in
all such measurements, an obvious question to ask is whether
or not we can think of this (anti-) correlation as a real property
of the two particle system independent of measurement.

Could it be that what is real about two systems in the singlet
state are
 not the local spin values,
 but merely the correlations between the two systems? Is quantum
mechanics about a world consisting not of objective values of
quantities but solely of objective correlations, of which
some are revealed in experiment? In other words, is there a
fundamental difference according to quantum mechanics as regards
the physical status of values of quantities and of correlations,
as for example Mermin \cite{mermin} seems to suggest?

 There is good reason to think that these questions should be
 answered in the positive, since a non-trivial theorem (which is true
 in quantum mechanics) points into this direction.
  The theorem (to be treated in the next section) shows that the global state of a composite quantum system
 can be completely determined by specifying
 correlations (joint probability distributions) when sufficient
 local measurements are performed on each subsystem.
 It thus suffices to consider only correlations when completely specifying
 the  state of the composite system.
  But can one also think of these correlations to be
  objective properties that pertain to the composite quantum system in question?
  As mentioned already in the case of the anti-correlation
  of the singlet state, one is tempted to think that this is indeed the case.
  However in this letter we will demonstrate that, however tempting,
  no such interpretation is possible and that
  these questions (as well as the questions
mentioned earlier) can thus not be answered in the positive.

Cabello \cite{cabello} and Jordan
\cite{jordan} give the same answer to similar questions using a Kochen-Specker \cite{kochenspecker}, Greenberger-Horne-Zeilinger (GHZ)
 \cite{ghz} or Hardy \cite{hardy}
argument.
Besides giving the Bell inequality version of the argument
(which in a sense completes the discussion because it was still lacking),
the advantage of the argument given here above these previous arguments,
is that it is more easily experimentally accessible using current
technology.
For this purpose, we explicitly present
a quantum state and the measurements that are to be performed in
order to test the inequality.

The structure of this paper
 is as follows.
In sec. \ref{qmcorr}
we will present an argument to the effect that quantum correlations
are real objective properties pertaining to composite quantum systems.
In the next two sections we will however show that this
  line of thought is in conflict with quantum mechanics itself.
To get such a conclusive result we need to be very formal and rigorous. In sec. \ref{bell} we will
therefore define our notion of correlation and derive a Bell inequality for correlations using a stochastic
hidden variable model under the assumption of local realism,
which formalises the idea of correlations as objective local properties.
 In sec. \ref{qmbell} we show that this inequality, when turned into it's
quantum mechanical form, is violated by quantum correlations. We
present a quantum state and a set of measurements that allow for
such a violation and furthermore show that it is the maximum
possible violation.
In sec. \ref{ontol} we apply this result to show that
entanglement cannot be considered ontologically robust
 when the quantum state is taken to be a complete description.
However, we argue that it nevertheless can be considered a resource
in quantum information theory to perform computational and
information-theoretic tasks.
In the last section, sec. \ref{conc},  we briefly
discuss the implications of our results, compare our argument
 to the ones given by Cabello \cite{cabello} and by Jordan
\cite{jordan} and return to the questions stated in the beginning of this introduction.

\section{DOES THE QUANTUM WORLD CONSIST OF CORRELATIONS?}\label{qmcorr}

In many important instances a system can be regarded as composed out
of separate subsystems.
In a physical theory that describes such composite
systems it can be asked whether one can assume that
the global state of the system
 can be completely determined by specifying
 correlations (joint probability distributions)
 when a sufficient number of local measurements
 are performed on each subsystem
 (note that here (and in the rest of the paper)
`local' is taken to be opposed to `global'
and thus not in the sense of spatial localisation\footnote{Local
thus refers to being confined to a subsystem of a larger system, without requiring the
subsystem itself to be localised (it can thus itself exist of
spatially separated parts).}).
  Barrett \cite{barrett} calls this the \emph{global state assumption}.
 Perhaps not surprisingly, the assumption holds for classical
 probability theory and for quantum mechanics on a complex Hilbert space.
 However, it need not be satisfied in an arbitrary theory,
 which shows that the theorem is non-trivial. For example, Wootters \cite{wootters}
 has shown that for quantum mechanics on a real Hilbert
 space the assumption does not hold because the correlations between subsystems do not suffice
 to build up the total state. By counting available degrees of freedom of
 the state of a composite system and of the states of its subsystems one
 can easily convince oneself that this is the case
  \footnote{J. Barrett (private communication) gives the following counting argument.
  A density matrix on a real Hilbert space with dimension $d$ has
  $N= (d^2-d)/2 +d=(1/2)d(d+1)$ parameters (without normalization),
  and a density matrix on a $d\otimes d$-dimensional Hilbert space
  has $(1/2)d^2(d^2 +1)$ parameters, which is too many because it is more than $N^2$.}.

Mermin has called the fact that in quantum mechanics the global state assumpion holds
\emph{sufficiency of subsystems correlations}, or the SSC theorem \cite{mermin2}. He phrases it as follows.  Given a system $\mathcal{S}=\mathcal{S}_1+\mathcal{S}_2$
with density matrix $W$,
then $W$ is completely determined by correlations
(joint probability distributions) Tr$[W(A_i\otimes B_j)]$ for an appropriate set
of observable pairs $\{A_i\},\{B_j\}$,
where $A_i=A_i\otimes \1$ is an observable for subsystem
$\mathcal{S}_1$ and $B_j=\1\otimes B_j$ is an observable for subsystem
$\mathcal{S}_2$.
The proof \footnote{Wootters has also independently proven this,
see \cite{wootters}.} relies on three facts:
Firstly, the mean values of \emph{all} observables for the entire system determine its state.
  Secondly, the set of all products over subsystems of subsystem observables
  (i.e., the set $\{ A_i\otimes B_j \}$) contains
 a basis for the algebra of \emph{all} such system-wide observables.
  Thirdly, the algorithm that supplies observables with their mean value is
linear on the algebra of observables.

As an example of the theorem, consider the well known singlet state written
as the one-dimensional projection operator
 \beq
\hat{P}_{singlet}= \frac{1}{4}(\1 -\sigma_z^1 \otimes\sigma_z^2
-\sigma_x^1 \otimes\sigma_x^2-\sigma_y^1 \otimes\sigma_y^2).
\eeq
The mean value of $\hat{P}_{singlet}$ is determined by the mean values
of the products of the  $x$, $y$ and $z$ components of the individual spins.
Since the mean value of this projector is 1 for the singlet,
the singlet state is thus determined by
the spin correlations in $x$, $y$ and $z$ direction having the value $-1$
(perfect anti-correlation). Perfect anti-correlation of \emph{any} three orthogonal components is
thus enough to ensure that the global state is the singlet state.
Thus correlations among all subsystems completely determine
the density matrix for the composite system they make up,
or in Mermin's words \cite{mermin2}:
``anything you can say in terms of quantum states can be translated into
a statement about subsystem correlations, i.e., about joint distributions."

It is tempting to think that because of
this theorem and because of the fact that Bell has shown that
a quantum state is not a prescription of local realistic values of
physical quantities,
that we can take a quantum state to
be nothing but the encapsulation of all the quantum correlations
present in the quantum system.
Indeed, the SSC theorem was used by N. David Mermin in 1998
to argue for the idea that correlations are physically real whereas
values of quantities are not \cite{mermin}
(although by now he has set these ideas aside\footnote{N.D. Mermin, personal communication.}).
Without wanting to claim that Mermin is committed to the issue we address next,
we explore if correlations between subsystems of an individual isolated composed system,
although determining the state of the total composite system,
can also be be considered to be real objective local properties of such a system.
That is, can one consider quantum correlations to be local realistic properties that
somehow (pre-)exist in the quantum state? Are correlations somehow atomic building blocks of the (quantum) world?

In the next two sections we will show that none of these questions
can be answered in the positive. We will be formal and rigorous and
follow the road paved by John Bell for us, but enlargen it
to not only include values of quantities but also correlations.

\section{A BELL INEQUALITY FOR CORRELATIONS}\label{bell}
\noindent Consider two spatially separated parties $I$ and $II$ which each have a
bi-partite system. Furthermore,
assume that each party determines the correlations of the
bi-partite system at his side. By correlations we here mean the
joint probability distributions $P^I_{AB}(ab)$ and $P^{II}_{CD}(cd)$,
where $A$ and $B$ are physical quantities each associated
to one of the subsystems in the bi-partite
system that party $I$ has, and where $a$ and $b$ denote the possible
values these quantities can obtain. The same holds for quantities
$C$, $D$ and possible outcomes $c$, $d$
but then for party $II$.
We now assume local realism for these correlations
in the following well-known way. Firstly, the correlations party $I$ finds are
determined by some hidden variable $\lambda \in \Lambda$ (with distribution
 $\rho(\lambda)$ and hidden variable space $\Lambda$).
 The same of course holds for $II$.
 Secondly, because of locality the correlations one party will obtain
 are for a given $\lambda$ statistically independent of the correlations
that the other party will find\footnote{
To be more specific, the doctrine of local realism leading to the Bell-inequalities
is the conjunction of the following three assumptions.
(i) \emph{Outcome independence}. The probability to obtain results
for the observables measured on one side is completely determined by
the experimental setup and the hidden
variable $\lambda$. There is no dependency
on the results obtained on the other side.
(ii) \emph{Parameter independence}. The probability to obtain results is only
locally determined, i.e., it is independent of the distant measurement devices.
(iii) \emph{Autonomy of the source}.
The hidden variable distribution $\rho(\lambda)$ of the source is
independent of the particularly chosen measurement setup.
The conjunction of these three lead to the result that the joint
probability distribution $P_{AB,CD}(ab,cd\,|\,\lambda)$ factorises:
 $P_{AB,CD}(ab,cd\,|\,\lambda)=P_{AB}^I(ab\,|\,\lambda)\,
P_{CD}^{II}(cd\,|\,\lambda)$. Thus conditioned on $\lambda$ the probabilities to obtain
outcome pairs $a$, $b$ and $c$, $d$ when measuring $A,B$ and $C,D$ are statistically independent.
}.
 Under these assumptions the joint
probability distribution
 over the four possible outcomes factorises which gives:
 \beq \label{factorisability}
P_{AB,CD}(ab,cd)= \int_{\Lambda}P_{AB}^I(ab\,|\,\lambda)\,
P_{CD}^{II}(cd\,|\,\lambda)\, \rho(\lambda)\,d\lambda.
 \eeq
Here we assume a so-called stochastic hidden variables model where the
hidden variables $\lambda$ determine only the joint probabilities
$P_{AB}^I(ab\,|\,\lambda)$, $P_{CD}^{II}(cd\,|\,\lambda)$ (i.e., correlations), and not the
values of the quantities themselves.
Furthermore, the joint probability $P_{AB}(ab\,|\,\lambda)$
itself need not factorise (the same of course holds for $P_{CD}(cd\,|\,\lambda)$.

Suppose now that we deal with dichotomic quantities $A,B,C,D$
with possible outcomes  $a,b,c,d \in \{-1,1\}$.
 The mean value of the product of two correlations is given by
\beq
E(AB,CD)= \sum_{a,b,c,d } abcd \,P_{AB,CD}(ab,cd).
\eeq
Then because of the factorisability of Eq.(\ref{factorisability})
we get the following
Bell inequality, in so-called CHSH form \cite{chsh},
\begin{eqnarray}\label{bellineq}
     \lefteqn{|E(AB,CD) +E(AB,(CD)')}\nonumber\\ & & +E((AB)',CD)-E((AB)',(CD)')|\leq 2.
\end{eqnarray}
Here $AB,\,(AB)'$ denote two sets of quantities that give rise to
two different joint probabilities (i.e., correlations) at party
$I$. Similarly for the set $CD$ and $(CD)'$ at party $II$.

This is the Bell inequality in terms of correlations that will be used in the next section.
Despite the resemblance between our inequality and the usual CHSH inequality,
they are fundamentally different because the latter
is in terms of subsystems quantities whereas the former is in terms of correlations and
does not assume anything about the values of subsystems quantities,
which the usual inequality does.

\section{QUANTUM CORRELATIONS ARE NOT LOCAL ELEMENTS OF REALITY}
\label{qmbell}
Consider a four-partite quantum system $\mathfrak{O}$ that consists of two
pairs of spin-$\frac{1}{2}$ particles where parties $I$ and $II$ each receive a single pair.
In this section we will provide an entangled state of the four-partite
quantum system $\mathfrak{O}$ and specific sets of bi-partite observables each performed
by parties $I$ and $II$ (that each have a bi-partite system)
with the following property: These observables give rise to
correlations (i.e., joint probability distributions) in the
bi-partite subsystems, which violate the Bell inequality
in terms of correlations of
the previous section (see Eq. (\ref{bellineq})). However, before doing that we have to give
the quantum mechanical version of this Bell inequality. We proceed as follows.

Firstly, by a quantum correlation we have in mind the
joint probability distribution of two (or more) self-adjoint operators,
 where each such operator corresponds to an observable of a subsystem.
Note that in case of independent observables this equals the distribution
of the product of the observables.
Such joint probability distributions are uniquely determined by
mean values of (sums of) products of projection operators
onto linear subspaces. In the remainder we can
therefore take quantum correlations to be the joint probability
distributions $P_{\hat{Q}\otimes \hat{R}}(qr\,|\, W)$ that are
obtained from products of projection operators $\hat{Q}$ and
$\hat{R}$ (with possible outcomes $q,r \in \{0,1\}$) for the
bi-partite state (density operator) $W$.
Of course $\hat{Q}$ and $\hat{R}$ must commute
in order for the joint probability distribution
to be well defined, but this is ensured since both
operators are defined for different subsystems
(with each their own Hilbert space) and are therefore commuting.

Secondly, we choose the set of hidden variables to
be the set of possible states on the Hilbert space $\H^{\mathfrak{O}}$ of
the four-partite system $\mathfrak{O}$, i.e.,
$\rho(\lambda)=\delta(\lambda -W_0)$, where $W_0$ is a
quantum state (i.e., density operator) on $\H^{\mathfrak{O}}$.
This is possible because quantum mechanics can be regarded as a
stochastic hidden variables theory where the hidden variable
is the quantum state. Upon assuming local realism, we then recover the same factorisability condition
of Eq. ({\ref{factorisability})
\beq \label{factorisabilityqm}
P_{\hat{A}\hat{B},\hat{C}\hat{D}}(ab,cd\,|\,W_0)=
P_{\hat{A}\hat{B}}^I(ab\,|\, W_I)\,P_{\hat{C}\hat{D}}^{II}(cd\,|\, W_{II}),
 \eeq
where $P^I_{\hat{A}\hat{B}}(ab\,|\,W_I)$ is the quantum mechanical
joint probability distribution to obtain outcomes $a$ an $b$
when measuring observables
$\hat{A}$ and $\hat{B}$ (each associated to a different subsystem)
on the bi-partite reduced state $W_I$, and the same holds
for the other joint probability distribution in the case of party $II$.
 This factorisability condition
finally gives us the quantum mechanical version of the Bell inequality of Eq. (\ref{bellineq})
encoding local realism for correlations:
\BEQ\label{bellineqqm}
 & & |E_{W_0}(\mathfrak{B})|=   |E_{W_0}(\hat{A}\hat{B},\hat{C}\hat{D}) +E_{W_0}(\hat{A}\hat{B},(\hat{C}\hat{D})')\nonumber \\
& & +E_{W_0}((\hat{A}\hat{B})',\hat{C}\hat{D})-E_{W_0}((\hat{A}\hat{B})',(\hat{C}\hat{D})'
)|\leq 2,
\EEQ
where $E_{W_0}(\hat{A}\hat{B},\hat{C}\hat{D})=
\rm{Tr} [W_0\,\hat{A}\otimes\hat{B}\otimes\hat{C}\otimes\hat{D}]$,
and $\mathfrak{B}$ is the so called Bell operator.

Now that we have the quantum version of the
Bell-inequality in terms of correlations, we will
provide an example of a violation of it.
Consider two sets of two dichotomic observables
represented by self-adjoint operators $\hat{a},~ \hat{a}'$ and
$\hat{b},~\hat{b}'$ for party $I$ and $II$ respectively. Each
observable acts on the subspace
$\H=\mathbb{C}^2\otimes\mathbb{C}^2$ of the bi-partite system held
by the respective party $I$ or $II$. These observables are chosen
to be dichotomous, i.e. to have possible outcomes in $\{-1,1\}$.
They are furthermore chosen to be
sums of projection
operators and thus give rise to unique joint probability
distributions on the set of quantum states. Measuring these
observables thus implies determining some quantum correlations.
For these observables  $\hat{a}$, $\hat{a}'$, $\hat{b}$ and $\hat{b}'$
the Bell operator $\mathfrak{B}$  on
$\H=\mathbb{C}^2\otimes\mathbb{C}^2\otimes\mathbb{C}^2\otimes\mathbb{C}^2$
becomes
$\mathfrak{B}=\hat{a}\otimes\hat{b}+\hat{a}\otimes\hat{b}'+\hat{a}'\otimes\hat{b}-
\hat{a}'\otimes\hat{b}'$.
The observables have the following form. Firstly, \beq\label{observable1}
\hat{a}=\hat{P}_{\psi^+}+\hat{P}_{\phi^+}-\hat{P}_{\psi^-}-\hat{P}_{\phi^-},
\eeq which is a sum of four projections onto the Bell basis
$\ket{\psi^\pm}=1/\sqrt{2} (\ket{\uparrow\downarrow} \pm
\ket{\downarrow\uparrow})$ and $\ket{\phi^\pm}=1/\sqrt{2}
(\ket{\uparrow\uparrow} \pm\ket{\downarrow\downarrow})$. Here the
states $\ket{\uparrow}$ and $\ket{\downarrow}$ are the spin-states
for ``up'' and ``down'' in the $z$-direction of a single particle
respectively, and together they form a basis for $\H=\mathbb{C}^2$.
Secondly, \beq
\hat{a}'=\hat{P}_{\ket{\uparrow\uparrow}}+\hat{P}_{\ket{\uparrow\downarrow}}-\hat{P}_{\ket{\downarrow\uparrow}}
-\hat{P}_{\ket{\downarrow\downarrow}}, \eeq where the projections
are onto the product states
$\ket{\uparrow\uparrow},~\ket{\uparrow\downarrow},~\ket{\downarrow\uparrow},~\ket{\downarrow\downarrow}$.
And finally, \beq \hat{b}=
\hat{P}_{\ket{\uparrow\uparrow}}+\hat{P}_{\ket{b+}}
-\hat{P}_{\ket{b-}}-\hat{P}_{\ket{\downarrow\downarrow}}, \eeq
\beq\label{observable4}
\hat{b}'=\hat{P}_{\ket{\downarrow\downarrow}}+\hat{P}_{\ket{b'+}}
-\hat{P}_{\ket{b'-}}-\hat{P}_{\ket{\uparrow\uparrow}}, \eeq
\noindent
where we have $\ket{b
\pm}=C^\pm(\ket{\uparrow\downarrow} + (1\pm \sqrt{2})
\ket{\downarrow\uparrow})$ and $\ket{b'
\pm}=C^\mp(\ket{\uparrow\downarrow} +(-1\pm \sqrt{2})
\ket{\downarrow\uparrow})$, with normalization coefficients
$C^\pm=(4\pm2\sqrt{2})^{-1/2}$
\footnote{
This particular choice of observables $\hat{a}$, $\hat{a}'$, $\hat{b}$, $\hat{b}'$ on
$\H=\mathbb{C}^2\otimes\mathbb{C}^2$ is motivated by a
well-known choice of single particle observables that allows for a
maximum violation of the original Bell-CHSH inequality when using the singlet state $\ket{\phi^-}=1/\sqrt{2}
(\ket{\uparrow\uparrow} -\ket{\downarrow\downarrow})$. This choice is
$\hat{a}=-\hat{\sigma_x},~\hat{a}'=\hat{\sigma_z},~
\hat{b}=1/\sqrt{2}(\hat{-\sigma_z} +\hat{\sigma_x}),~
\hat{b}'=1/\sqrt{2}(\hat{\sigma_z} +\hat{\sigma_x})$
all on $\H=\mathbb{C}^2$.
The analogy can be seen by noting that in this latter choice the (unnormalized) eigenvectors of
$\hat{a}$ are $\ket{\uparrow}+\ket{\downarrow},~\ket{\uparrow}-\ket{\downarrow}$,
of $\hat{a}'$ they are $\ket{\uparrow},~ \ket{\downarrow}$,
of $\hat{b}$ they are $\ket{\uparrow} +(1+\sqrt{2})\ket{\downarrow},~
\ket{\uparrow} +(1-\sqrt{2})\ket{\downarrow}$
and finally of $\hat{b}'$ they are $\ket{\uparrow} +(-1+\sqrt{2})\ket{\downarrow},~
\ket{\uparrow} +(-1-\sqrt{2})\ket{\downarrow}$.
}.

Consider now the four particle entangled pure
state \beq\label{state} \ket{\Psi}=\frac{1}{\sqrt{2}}
(\ket{\uparrow\downarrow\uparrow\downarrow}-\ket{\downarrow\uparrow\downarrow\uparrow}).
\eeq The mean value of the Bell operator $\mathfrak{B}$ for
the above choice of $\hat{a}$, $\hat{a}'$, $\hat{b}$, $\hat{b}'$
in the state $\ket{\Psi}$ is equal to
\beq
|{\rm Tr} [\mathfrak{B}\ket{\Psi}\bra{\Psi}] |=2\sqrt{2}.
\eeq
\noindent This gives us a violation of the Bell
inequality of Eq. (\ref{bellineqqm}) by a
factor of $\sqrt{2}$. This violation proves that
quantum correlations cannot be considered to be local elements of reality.

The violation is the maximum value because Cirel'son's inequality
\cite{cirelson} (i.e., $|{\rm Tr} [\mathfrak{B}W] |\leq
2\sqrt{2}$ for all quantum states $W$) must hold for all
dichotomic observables $\hat{a},\hat{a}',\hat{b},\hat{b}'$ on
$\H=\mathbb{C}^2\otimes \mathbb{C}^2$ (possible outcomes in $\{-1,1\}$). One can easily see this
because for $\hat{a},\hat{a}',\hat{b},\hat{b}'$
we have that
$\hat{a}^2= \hat{a}'^2=\hat{b}^2=\hat{b}'^2=\1$,
 and it thus follows that the proof of Landau \cite{landau} of  Cirel'son's inequality
 goes through.

\section{ENTANGLEMENT IS NOT ONTOLOGICALLY ROBUST}\label{ontol}

Entanglement is the fact that certain quantum states of a composite system exist
that are not convex sums of product states\footnote{For completeness, a bipartite state $W$
is \emph{entangled} iff $W \neq \sum_i p_i W_{i}^I\otimes W^{II}_i$,
where $W_{i}^I$ and $W^{II}_i$ are arbitrary states of the two subsystems
 and $\forall ~p_i>0$, $\sum_i p_i =1$.
 A state that is not entangled is called \emph{separable}.
 For the definition of
multipartite entanglement see \cite{seevinck}.
}.
It gives rise to a special kind of quantum correlations, called
 non-classical correlations, which in the pure state case allow one to violate a Bell inequality
(for mixed states it is not always the case that entanglement
implies violation of a Bell inequality \cite{werner}).  Furthermore, these correlations can also be used to
perform exceptional quantum information and computation tasks.

 The SSC theorem of section \ref{qmcorr} tells us that
 entanglement can be completely characterised by
the above quantum correlations that it gives rise to.
 Therefore the result of the previous section also applies to entanglement.
Then, if one considers the quantum state description to be complete,
entanglement cannot be viewed as ontologically robust
in the sense of being a local objective
 property pertaining to some composite system.
  If one would do so nevertheless, one can construct a composite system
 that contains as a subsystem
 the entanglement (i.e. the entangled system) in question and
 which would allow for a violation of the
 Bell inequality of Eq. (\ref{bellineqqm}). This implies
 (contra the assumption) that the entanglement cannot be regarded
 in a local realistic way, which we take to be
 a necessary condition for ontological robustness.

It is possible that one thinks that the requirement of local
 realism is too strong a requirement for ontological robustness.
 However, that one cannot think of entanglement as a property
wich has some ontological robustness
 can already be seen using the following weaker requirement:
 anything which is ontologically robust can, without interaction, not
be mixed away, nor swapped to another object, nor
flowed irretrievably away into some environment.
 Precisely these features are possible in the case of entanglement
 and thus even the weaker requirement for ontological robustness does not hold.

 These features show up at the level of quantum states when considering a quantum system in
conjunction with other quantum systems: entanglement can (i) be created in previously
 uninteracting particles using swapping,
(ii) be mixed away and (iii) flow into some environment upon mixing, all without interaction
 between the subsystems in question.
It is this latter point, the fact that no interaction
 is necessary in these processes, that one cannot think of entanglement
 as ontologically robust.


To see that the above weaker requirement
   for ontological robustness of entanglement does
 not hold consider the following examples
 of the three above mentioned features.

  (i) Consider two maximally entangled pairs (e.g. two singlets) that are created
 at spacelike separation, where from each pair a particle is emitted
 such that these two meet and the other particle of each pair is emitted such
 that they fly away in opposite directions. Conditional on a suitable joint measurement performed on
 the pair of particles that will meet (a so called Bell-state measurement)
  the state of the remaining two particles, although they have never
  previously interacted, will be `thrown'
  into a maximally entangled state. The entanglement is swapped \cite{swap}.

 (ii) Equally mixing the two maximally entangled Bell states
$\ket{\psi^\pm}$
 gives the \emph{separable} mixed state
 \beq \label{mix}W=
 \frac{1}{2}(P_{\ket{\psi^+}} +P_{\ket{\psi^-}}
) =
 \frac{1}{2}(P_{\ket{\uparrow\downarrow}} +P_{\ket{\downarrow\uparrow}}).\eeq
   The entanglement is thus mixed away, without any necessary interaction between
   the subsystems.

  (iii) Equally mixing the following two states of three spin 1/2 particles,
where particles $2$ and $3$ are entangled in both states,
 \BEQ\nonumber\ket{\psi}&=&\frac{1}{2}( \ket{\uparrow}^{(1)}\otimes\ket{\psi^{-}}^{(23)})\\
  \ket{\phi}&=&\frac{1}{2}(\ket{\downarrow}^{(1)}\otimes\ket{\psi^{+}}^{(23)}),
  \EEQ
gives the state
\beq W= \frac{1}{2}(P_{\uparrow}^{(1)}\otimes P_{\psi^{-}}^{(23)} +
 P_{\downarrow}^{(1)}\otimes P_{\psi^{+}}^{(23)}). \eeq
This three-particle state is two-particles entangled although it has \emph{no}
two particle subsystem whose (reduced) state is entangled \cite{seevinck}.
The bipartite entanglement has thus irretrievably flowed into the three particle state,
again without any necessary interaction between the subsystems.

Does this lack of ontological robustness of entanglement question
the popular idea of entanglement as a
resource for quantum information and computation tasks?
We think it does not. Quantum information theory is precisely a theory devised to deal with
the surprising characteristics of entanglement such as the
ontologically non-robustness here advocated
(and many other features, such as for example teleportation).
Entanglement is taken to a be specific type of correlation
that can be used as a resource for encoding and manipulating (quantum)
bits of information. For that purpose the ontological status of the information or of that which
bears the information does not matter. The only thing that matters is that entanglement can be manipulated
 according to certain rules that allow for interesting information-theoretic and computational tasks.

To conclude this section we should mention that Timpson and Brown \cite{timpson} do argue for the ontological
robustness of entanglement in the mixing case (ii) above
by introducing ontological relevance
to the preparation procedure of a quantum state,
which supposedly can always be captured in the full quantum mechanical description.
 They introduce the dictinction between
 `improper' and `proper separability', which is analogous
 to the well known distinction between proper and improper mixtures,
 to argue that one can retain
 an ontologically robust notion of entanglement.
They thus call the separable mixed state of Eq. (\ref{mix}) improperly separable
because the entanglement in the mixture becomes hidden on mixing
(i.e, it disappears), although there are
some extra facts of the matter
that tell that the separable state is in fact
composed out of an ensemble of entangled states.
We agree that ``there need be no mystery at the conceptual level
over the disappearance" (sec 2. in ref. \cite{timpson}),
and the introduced distinction between proper and improper
separability indeed shows this. However, we are not convinced
that their analysis of the improperly separable states allows for ontological robustness of entanglement.

Timpson and Brown argue that using the extra facts of the matter an observer
 is able to perform
a place selection procedure that would allow the
ensemble to be separated out into the original
statistically distinct sub-ensembles
[i.e., into the entangled states].
According to them it is always the case that such
 a procedure exists \emph{in principle}:
``all that is required is access to these further facts"
(sec. 1 in ref. \cite{timpson}).

However, we believe that it is very well
conceivable that according to quantum mechanics
we do not in principle always have access to these extra
facts.
Perhaps the interactions between the object systems involved in
the preparation procedure and the environment are such that the
observer cannot become correlated to both
the extra facts and the objects states
in the right way for the facts to be accessable,
or the interactions could be such that no classical record of
the extra facts could possibly be left in the environment.
To put it differently,
although we agree that in the case of improper separability
one can uphold an ignorence interpretation of the state in question
and that furthermore the ignorance is in principle about some extra
facts of the matter, we do not agree that it is certain that the ignorance about these extra facts
of the matter can be removed by the observer
in accordance with the dynamics of quantum mechanics in all
 conceivable preparation procedures.
 The issue thus awaits a (dis)proof of principle,
 which the author wants to undertake in the near future.

\section{DISCUSSION AND CONCLUSION}\label{conc}
\noindent The Bell inequality violation of section \ref{qmbell}
tells us that despite the fact that a quantum state of a composite
system is determined by the correlations between
each of its possible subsystems, one cannot conclude that they are determined (by
the quantum state) in a local way.
Just like
values of quantities correlations cannot be used to build up a world
consisting out of some local realistic structure.
We have that \emph{mathematically} quantum correlations determine the quantum state,
but \emph{ontologically} they cannot be considered to be
atomic building blocks of the world.
Furthermore, although entanglement
is taken to be a resource in quantum information theory,
we have argued that it cannot be considered ontologically robust
because it is not a local objective property
and furthermore that without interaction it can be mixed away, swapped to another object,
and flowed irretrievably into some environment.

The Bell inequality argument of section \ref{qmbell} was inspired by the work of
Cabello \cite{cabello} and Jordan \cite{jordan}
who obtain almost exactly the same conclusion,
although by different arguments.
The argument of Cabello differs the most from ours because
he uses a different conception of what a quantum correlation is. His argument
speaks of types of correlations which are associated with eigenvalues of
product observables. We believe this notion to be less general
than our notion of quantum correlation which only takes
joint probability distributions to be correlations.
  Jordan's argument, in contrast, does in effect use the same notion of quantum
 correlation as we do. He considers mean values of products
 of observables and since these are determined by mean values of
(sums of) products of projection operators he restricts himself to the latter.
He thus uses the same notion as we do because the latter determine all
joint probability distributions.

However, Cabello and Jordan both need perfect
correlations for their argument to work because the state dependent
GHZ- or Hardy argument they use (Cabello uses both, Jordan only the latter
argument) need such strong correlations.
Our Bell inequality argument does not rely on this
specific type of correlation because statistical correlations already suffice
to violate the Bell inequality here presented.
We therefore believe our argument
has an advantage over the one used by Cabello and Jordan,
because it is more amenable to experimental implementation.

     In fact, the Bell-inequality argument here presented can be readily
implemented using current experimental technology.
Indeed, it is already possible to create fully four-particle-entangled states
\cite{4deeltjesexperiment}
and measurement of the four observables $\hat{a}$, $\hat{a}'$, $\hat{b}$, $\hat{b}'$ of
Eq.(\ref{observable1})-(\ref{observable4}) seems not to be
problematic since they are sums of ordinary projections.
Furthermore, as said before, there is no need to produce perfect correlations;
statistical correlations will suffice. We therefore hope that in
the near future experiments testing our argument will be carried
out.

Lastly, returning to the questions stated in the introduction,
in so far as Mermin in his \cite{mermin} is committed to
take correlations (as we have defined them here) to be locally real
(which we think he is), his tentative interpretation is at odds
with predictions of quantum mechanics and would allow,
in view of the argument given here, for an experimental verdict.

\subsection*{Acknowledgements}\noindent
I would like to thank Jos Uffink for valuable discussions,
Sven Aerts and N. David Mermin for helpful comments on an earlier draft
 and Harvey Brown for stimulating comments on this topic,
 even though the latter go back many years.



\begin{thebibliography}{99}
\bibitem{bell}J.S. Bell, " On the Einstein-Podolsky-Rosen paradox," \emph{Physics} {\bf 1}, 195 (1964).
\bibitem{mermin}\label{merminlabel}N.D. Mermin, in a series of papers, tried to defend this
 fundamental interpretational difference between values of quantities and correlations.
 N.D. Mermin, "What is quantum mechanics trying to tell us?," \emph{Am. J. Phys.} {\bf 66}, 753 (1998);
N.D. Mermin, "What do these correlations know about reality? Nonlocality and the absurd," \emph{Found. Phys.} {\bf 29}, 571 (1999);
N.D. Mermin, "The Ithaca interpretation of quantum mechanics," \emph{Pramana} {\bf 51}, 549 (1998).
\bibitem{cabello}A. Cabello, "Quantum correlations are not contained in the initial state," \emph{Phys. Rev. A} {\bf 60}, 877 (1999).
\bibitem{jordan}T.F. Jordan, "Quantum correlations violate Einstein-Podolsky-Rosen assumptions," \emph{Phys. Rev. A} {\bf 60}, 2726 (1999).
\bibitem{kochenspecker} S. Kochen and E.P. Specker, "The problem of hidden variables
in quantum mechanics,"
\emph{J. Math. Mech.} {\bf 17}, 59 (1967).
\bibitem{ghz} D.M. Greenberger, M.A.Horne and A. Zeilinger, "Going beyond Bell's theorem," in {\emph{Bell's theorem,
quantum theory and conceptions of the universe}},  M. Kafatos (Ed.), (Kluwer Academic, Dordrecht, 1989), p. 69;
D.M. Greenberger, M.A.Horne, A. Shimony, and A. Zeilinger, "Bell's theorem without inequalities,"
 \emph{Am. J. Phys.} {\bf 58}, 1131 (1990).
\bibitem{hardy} L. Hardy, "Nonlocality for two particles without inequalities for almost all states," \emph{Phys. Rev. Lett.} {\bf 71}, 1665 (1993).
\bibitem{barrett} J. Barrett, "Information processing in non-signalling theories," Los Alamos e-print archive,
{ http://arxiv.org/abs/quant-ph/0508211}.
\bibitem{wootters} W.K. Wootters,"Local accessibility of quantum states," in {\emph{Complexity, Entropy and the Physics of Information}},
 W.H. Zurek (Ed.), (Addison-Wesley, Boston, 1990), p. 39.
\bibitem{mermin2} See the first paper by N.D. Mermin in \cite{mermin}.
\bibitem{chsh}J.F. Clauser, W.A. Horne, A. Shimony,  and R.A. Holt,
"Proposed experiment to test local hidden-variable theories,"
\emph{Phys. Rev. Lett.} {\bf 23}, 880 (1969).
\bibitem{cirelson} B.S. Cirelson, "Quantum generalizations of Bell's inequality," \emph{Lett. Math. Phys.} {\bf 4}, 93 (1980).
\bibitem{landau}L.J. Landau, "On the violation of Bell's inequality in quantum theory," \emph{Phys. Lett. A.} {\bf 120}, 54 (1987).
\bibitem{werner}R.F. Werner, "Quantum states with Einstein-Podolsky-Rosen
correlations admitting a hidden-variable model," \emph{Phys. Rev. A} {\bf 40}, 4277 (1989).
\bibitem{seevinck} M. Seevinck and J. Uffink, "Sufficient conditions for three-particle
entanglement and their tests in recent experiments," \emph{Phys. Rev. A} {\bf 65}, 012107 (2001).
\bibitem{swap} M. \.Zukowski, A. Zeilinger, M.A. Horne and A.K. Ekert,
"`Event-ready-detectors' Bell experiment via entanglement swapping,"
\emph{Phys. Rev. Lett.} {\bf 71}, 4287 (1993).
\bibitem{timpson} C.G. Timpson, H.R. Brown, "Proper and improper separability," \emph{Int. J. Quant. Inf.} {\bf 3}, 679 (2005).
\bibitem{4deeltjesexperiment}C.A. Sackett, \emph{et al.}, "Experimental entanglement of four particles,"
  \emph{Nature} {\bf 404}, 256 (2000); Z. Zhao, \emph{et al.}, "Experimental violation
  of local realism by four-photon Greenberger-Horne-Zeilinger entanglement,"
   \emph{Phys. Rev. Lett.} {\bf 91}, 180401 (2003);
     Z. Zhao, \emph{et al.}, "Experimental demonstration of five-photon entanglement and open-destination teleportation,"
\emph{Nature (London)} {\bf 430}, 54 (2004);
M. Eibl, \emph{et al.}, "Experimental observation of four-photon entanglement from parametric
down-conversion,"
 \emph{Phys. Rev. Lett.} {\bf 90}, 200403 (2003).
\end{thebibliography}
\end{document}